\documentclass[twocolumn,secnumarabic,amssymb, nobibnotes, aps, prd]{revtex4}

\usepackage{graphicx}
\usepackage{dcolumn}
\usepackage{bm}
\usepackage{color}
\begin{document}


\title{Non-linear transport phenomena in a two-subband system}

\author{S. Wiedmann,$^{1,2,3}$ G. M. Gusev,$^4$ O. E. Raichev,$^5$ A. K. Bakarov,$^6$ and J. C. Portal$^{2,3}$}
\affiliation{$^1$Radboud University Nijmegen, Institute for Molecules and Materials, High Field Magnet Laboratory, Toernooiveld 7, 6525 ED Nijmegen, the Netherlands} 
\affiliation{$^2$Laboratoire National des Champs Magn\'{e}tiques Intenses, CNRS-UJF-UPS-INSA, 38042 Grenoble, France} 
\affiliation{$^3$INSA Toulouse, 31077 Toulouse Cedex 4, France} 
\affiliation{$^4$Instituto de F\'{\i}sica da Universidade de S\~ao Paulo, CP 66318, S\~ao Paulo, SP,Brazil}
\affiliation{$^5$Institute of Semiconductor Physics, NAS of Ukraine, Prospekt Nauki 41, 03028, Kiev, Ukraine} 
\affiliation{$^6$Institute of Semiconductor Physics, Novosibirsk 630090, Russia}

\date{\today}

\begin{abstract}
We study non-linear transport phenomena in a high-mobility bilayer system with two closely 
spaced populated electronic subbands in a perpendicular magnetic field. For a moderate direct 
current excitation, we observe zero-differential-resistance states with a 
characteristic $1/B$ periodicity. We investigate, both experimentally and theoretically, the 
Hall field-induced resistance oscillations which modulate the high-frequency magneto-intersubband 
oscillations in our system if we increase the current. We also observe and describe the 
influence of direct current on the magnetoresistance in the presence of microwave irradiation. 
\end{abstract}


\maketitle

\section{Introduction}

Nonequilibrium magnetotransport in two-dimensional electron systems (2DES) at large filling 
factors is a subject of intense interest. Experimental and theoretical work includes studies of the 
steady-state response of 2DESs under ac excitation in the microwave (MW) range and under elevated direct 
current (dc). In both cases, there appears an oscillating magnetoresistance due to scattering-assisted 
transitions of electrons between different Landau levels (LLs). At low temperatures, when the 
elastic scattering dominates, such transitions may occur due to coupling of LLs either by the 
MW field, owing to absorption and emission of the radiation quanta $\hbar \omega$, or by the 
Hall field $E_{dc}$ which tilts the LLs. Consequently, the microwave-induced resistance oscillations 
(MIROs) \cite{1,2,3,4} are governed by the ratio $\epsilon_{ac}=\omega/\omega_{c}$ where $\omega_{c}=eB/m$ 
is the cyclotron frequency, $B$ is the magnetic field strength, and $m$ is the effective mass 
of electrons. Next, the Hall field-induced resistance oscillations (HIROs) \cite{5,6,7} are governed 
by the ratio $\epsilon_{dc}=2R_ceE_{dc}/\hbar\omega_{c}$, where $R_c=v_F/\omega_{c}$ is the 
classical cyclotron radius and $v_{F}$ is the Fermi velocity. The Hall field is defined as 
$E_{dc}=\rho_{H}j$, where $\rho_H = m \omega_c/e^2 n_s$ is the Hall resistivity, $n_s$ is the 
electron density and $j=I_{dc}/w$ is the current density with $I_{dc}$ being the applied 
current and $w$ the sample width. 

The amplitude of MIROs can increase dramatically with increasing MW power. In the 
samples with high electron mobility, elevated MW power transforms the minima of these 
oscillations into intervals of zero dissipative resistance. These are referred 
to as zero-resistance states (ZRSs) \cite{2,3,4}. The amplitude of HIROs is considerably 
smaller because these oscillations rely upon the large-angle scattering between field-tilted 
LLs, which requires the presence of short-range scattering potential. However, in the regime 
$\epsilon_{dc} < 1$, corresponding to the first minimum of the HIROs, the Hall field creates a 
strongly nonequilibrium electron distribution within the LLs, which causes a considerable decrease 
of the resistance \cite{8}. This may give rise to a phenomenon of vanishing differential 
resistance in high-mobility 2DESs \cite{9}; the resulting states are called zero-differential-resistance 
states (ZdRS). The ZDRSs can also emerge from the Shubnikov-de Haas (SdH) oscillation 
maxima \cite{10}. The ZDRS phenomenon in the dc response resembles the ZRSs in the 
MW-induced response and can be explained with similar reasons, in terms of the 
negative differential resistivity, which leads to an instability of the homogeneous current picture and 
spontaneous breaking of the sample into current domains \cite{11}. As the instability 
is of macroscopic origin, it exists regardless of the details of the microscopic mechanisms of 
non-equilibrium magnetoresistance. However, these mechanisms determine the region of the 
magnetic fields where the ZRSs or ZDRS are observed.  

Furthermore, an interesting situation takes place when a 2DES is subjected to both dc and 
ac excitations so that both sources of nonlinearity are applied together. Experimental studies 
\cite{12,13} have shown that the resulting oscillations depend on $\epsilon_{ac}$, and $\epsilon_{dc}$, 
as well as on the combination of these parameters, $\epsilon_{ac} \pm \epsilon_{dc}$. The component 
of the magnetoresistance oscillating with $\epsilon_{ac} \pm \epsilon_{dc}$ can be described as an 
interference between MIROs and HIROs. A continuous increase of the current transforms the oscillation 
maxima into minima (and vice versa) and may lead to destruction of ZRSs and to reappearance of ZRSs 
in a different region of magnetic fields.  

A unified physical picture of the nonequilibrium magnetotransport phenomena described 
above is currently under development. Earlier theoretical works were focused on the influence 
of MWs on the probabilities of elastic scattering of electrons in a system of LLs \cite{14,15,
16}, which contribute to the ``displacement" mechanism of MW-induced magnetoresistance. 
Later, the influence of both MWs and dc excitation on the electron distribution function was 
found to be very important \cite{17}. The deviation of this function from equilibrium 
is proportional to the inelastic electron-electron scattering time $\tau_{in}$, which rapidly 
increases with decreasing temperature. Thus, the corresponding ``inelastic" mechanism of 
non-linear magnetoresistance proves to be the most important one in the description of 
MW-induced magnetoresistance at low temperatures, which is also confirmed experimentally 
\cite{18}. A comprehensive consideration of possible mechanisms of MW-induced magnetoresistance 
and of their interplay at elevated radiation power and for different kinds of elastic disorder
potentials has been presented in Refs. \cite{19} and \cite{20}. A theory of dc magnetoresistance 
comprising both displacement and inelastic mechanisms and describing the response in a
wide interval of applied currents (including the region of HIROs) has also been developed \cite{21}. 
The interplay of both dc and ac excitations is studied in Refs. \cite{22} and \cite{23}, and the 
theoretical results confirm the basic properties of the oscillating magnetoresistance 
investigated in the experiments \cite{12} and \cite{13}.

The nonequilibrium magnetotransport phenomena acquire qualitatively distinct features 
in systems with two populated electronic subbands, because of the possibility of electron 
transitions between the two sets of Landau levels associated with these subbands. As a result, 
even in the linear regime, such systems show magneto-intersubband (MIS) oscillations 
governed by the ratio $\Delta/\hbar \omega_c$ \cite{24}. The subband separation energy 
$\Delta$ can be a few meV in double quantum wells so that the MIS oscillation 
period is comparable to the periods of MIROs and HIROs. Under MW excitation, the 
oscillating magnetoresistance is determined by an interference of MIROs and MIS 
oscillations \cite{25,26}. A theory based on consideration of the inelastic mechanism 
of magnetoresistance in a two-subband system satisfactorily explains the observed oscillating 
resistivity as well as its temperature dependence in the low-temperature region \cite{25}, 
\cite{27}, \cite{28}, \cite{29}. Similar to single-subband systems, high-mobility two-subband systems 
demonstrate the ZRS phenomenon \cite{28}. The non-linear response to moderate dc excitation 
\cite{30}, \cite{31} shows a flip (inversion) of MIS oscillations. This behavior signifies 
an inversion of the quantum contribution to resistivity as a consequence of nonequilibrium 
electron distribution, in agreement with the theory of Ref. \cite{17}. The dependence of 
the characteristic magnetic field (where the flip occurs) on the current density is in a 
quantitative agreement with the theory \cite{31}. The response to elevated dc excitation 
\cite{32} shows the interference of MIS oscillations with HIROs. Finally, it was recently 
found \cite{33} that high-mobility two-subband systems develop several regions of ZDRSs around 
the peaks of inverted MIS oscillations. 

In spite of those recent advances in studying magnetotransport in two-subband systems, some
problems are still poorly investigated. In particular, neither a systematic experimental 
study of HIROs nor their theoretical description has been presented. Preliminary studies of 
ZDRSs \cite{33} are limited to low temperatures and low dc currents. The behavior of two-subband 
systems under the combined action of dc and ac excitations has not been studied. In the present work, 
we intend to fill these gaps. We undertake studies of non-equilibrium magnetotransport 
in a high-mobility bilayer electron system formed by a wide quantum well (WQW). The high 
sample quality enables us to carry out a systematic experimental investigation of dc transport 
and to describe both ZdRS and HIROs which appear at elevated currents. We also show and discuss the 
influence of a dc electric field on MW-induced ZRSs at a chosen frequency of 143~GHz. A supporting 
theoretical consideration, based upon a generalization of existing theories to the two-subband 
case, reasonably explains the different regimes of non-linear magnetotransport. 

The paper is organized as follows. Section II briefly presents experimental details. In Sec. III we 
describe the magnetoresistance measured in a wide range of applied dc excitation and compare our 
data in the regime of HIROs with the theoretical calculations. In Sec. IV we analyze the influence 
of moderate and strong dc excitations on the magnetoresistance under MW irradiation. Concluding 
remarks are given in the last section.

\section{Experimental setup}

For our experiments, we use high-quality WQW's with a well width of 45~nm, 
high electron density $n_{s}\simeq 9.2\times10^{11}$~cm$^{-2}$, and a mobility of 
$\mu~\simeq 1.9 \times 10^{6}$~cm$^{2}$/V s after a brief illumination with a red 
light-emitting diode. Several samples in Hall bar geometry (length $l$ $\times$ width 
$w$) of (500 $\times$ 200) $\mu$m$^{2}$ have been studied. The electrons in our system
occupy the two lowest subbands with a subband separation of $\Delta=1.4$~meV,
extracted from the periodicity of low-field MIS oscillations. Measurements have been 
performed between 1.4 and 4.2~K in a cryostat with a variable-temperature insert. For 
experiments under MW irradiation, a waveguide is employed to deliver linearly 
polarized MW radiation down to the sample. We record longitudinal resistance using 
a current of 1~$\mu$A at a frequency of 13~Hz in the linear regime. Direct current 
$I_{dc}$ was applied simultaneously through the same current leads to measure the 
differential resistance $r_{xx}\equiv dV_{xx}/dI_{dc}$. In the next section we describe 
the measurements of the resistance under dc excitation and its theoretical analysis.

\begin{figure}[ht]
\includegraphics[width=9cm]{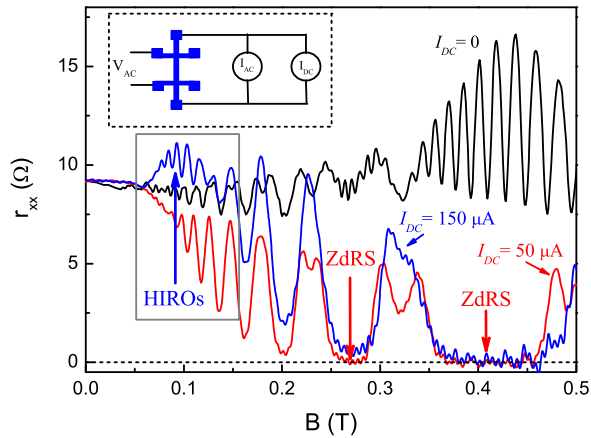}
\caption{\label{fig1} (Color online) Differential resistance $r_{xx}$ as a function of
the magnetic field at 1.4~K. ZdRS are present for high DC current, see arrows. In 
addition, HIRO's appear for $I_{dc}$=150~$\mu$A (see the box) at low magnetic fields. 
Inset: measurement setup for differential resistance.}
\end{figure}

\section{Magnetoresistance under dc excitation}

In Fig. \ref{fig1} we present the differential resistance $r_{xx}$ as a function of the 
magnetic field for $I_{dc}$=0, $I_{dc}$=50~$\mu$A, and $I_{DC}$=150~$\mu$A. For 
$I_{dc}$=0, the magnetoresistance exhibits well-developed MIS oscillations and thereby 
confirms the existence of two populated subbands \cite{24}. It should be noticed that 
the ac current of 1~$\mu$A is already high enough to invert MIS oscillations for 
$B < 0.2$~T. At 1.4~K, we also observe SdH oscillations which are superimposed on the MIS 
oscillations for $B>0.25$~T. We first apply a moderate current of 50~$\mu$A and find 
that the MIS oscillations are inverted in the whole range of magnetic fields. Around 0.27 and 
0.4~T, the inverted MIS peaks evolve into wide regions of ZDRSs. We also find a splitting 
of the oscillation maxima, e.g., at 0.23 and 0.32~T similar to that in dc-driven non-linear 
magnetotransport in double quantum wells using a high current \cite{31}. A further 
increase in the dc electric field, see trace for $I_{dc}$=150~$\mu$A, leads to 
vanishing ZDRSs at $B$=0.27~T and to disappearance of the splitting. Apart from that, 
the most intriguing features appear around $B$=0.1~T, where the inverted MIS peaks 
are transformed into enhanced ones. Other MIS peaks, e.g. at 0.2~T, remain inverted. 
We identify the feature around 0.1~T with the first HIRO peak. Indeed, if $I_{dc}$=150~$\mu$A the 
condition $\epsilon_{dc}=1$, which corresponds to the maximum of the first HIRO peak, occurs 
at $B \simeq 0.07$ T, not far from the point $B \simeq 0.09$ T where we actually observe 
the maximum. The shift of the observed maximum to higher fields occurs because the ocsillating 
contribution to resistivity is exponentially suppressed by the disorder in the region 
of low magnetic fields. 

\begin{figure}[ht]
\includegraphics[width=9cm]{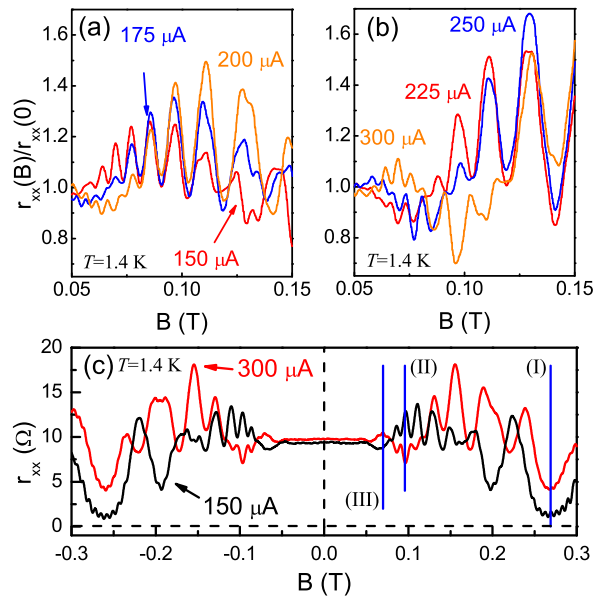}
\caption{\label{fig2} (Color online) (a), (b) Normalized differential resistance 
$r_{xx}(B)/r_{xx}(0)$ as a function of the magnetic field for several chosen dc 
currents showing HIRO's for elevated $I_{dc}$. (c) Differential resistance $r_{xx}$ for 
150 and 300~$\mu$A. Several MIS oscillations (fixed $B$) are marked with (blue) lines and 
named from (I) to (III). No SdH oscillations are present in the chosen range of magnetic fields.}
\end{figure}

\begin{figure}[ht]
\includegraphics[width=9cm]{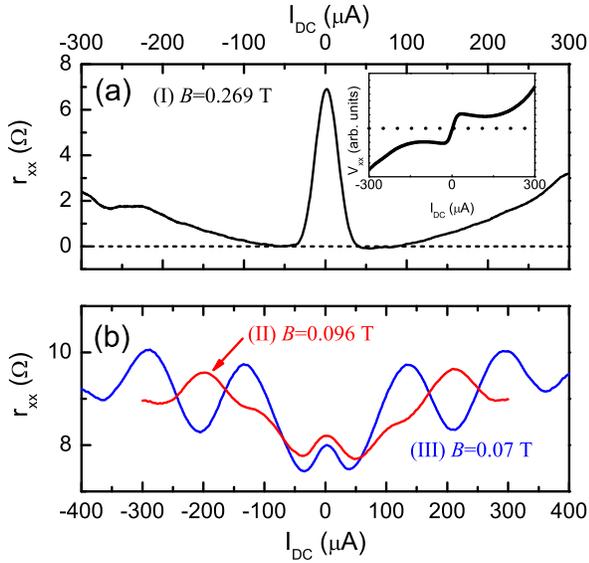}
\caption{\label{fig3} (Color online) Differential resistance as a function of the
applied DC current $I_{dc}$ for the inverted MIS peak which develops into a ZdRS (a)
and for two chosen MIS peaks II ($B$=0.096~T) and III ($B$=0.07~T from -400 to 400~ $\mu$A) 
at lower magnetic fields (b).}
\end{figure}

We now focus on HIRO's at elevated currents. Figures \ref{fig2} (a) and \ref{fig2} (b)
give an overview of magnetoresistance under dc excitation. In this range of magnetic
field from 0.05 to 0.15~T and at a temperature of 1.4~K, MIS oscillations are strongly
influenced by a dc electric field. The SdH oscillations, which occur at higher magnetic
fields are damped due to electron heating. 
For $I_{dc}$ between 150 and 
200~$\mu$A, we first see that all MIS oscillations that were inverted for 50~$\mu$A, 
are now strongly enhanced at $B > 0.07$~T. In addition, the features around 0.06~T in 
Fig. \ref{fig2}(a) become inverted once again with increasing dc excitation. A further 
increase in the dc excitation from 225 to 300~$\mu$A modifies the differential 
magnetoresistance in a similar way, by modulating the MIS oscillations according 
to the HIRO periodicity. This behavior is also illustrated in Fig. \ref{fig2}(c), 
where we plot $r_{xx}$ for both negative and positive $B$. 
In order to study MIS oscillation behavior with increasing electric field, we mark 
three MIS peaks (I) to (III), which we investigate in detail in Fig. \ref{fig3}.

In Fig. \ref{fig3}(a) we first show $r_{xx}$ as a function of $I_{dc}$ at $B$=0.269~T 
(as well as the corresponding voltage $V_{xx}$) where we observe a ZDRS. This state is 
created at $I_{dc}^{min}\simeq$40~$\mu$A and is observed until a maximal current 
$I_{dc}^{max}\simeq$80~$\mu$A. Figure \ref{fig3}(b) presents differential
resistance as a function of $I_{dc}$ for MIS peaks (II) and (III) with the 
corresponding magnetic fields indicated. For 0.07 T (III), the resistivity 
shows oscillations which are clearly periodic with the current, and the period 
corresponds to that of HIROs.

\begin{figure}[ht]
\includegraphics[width=9cm]{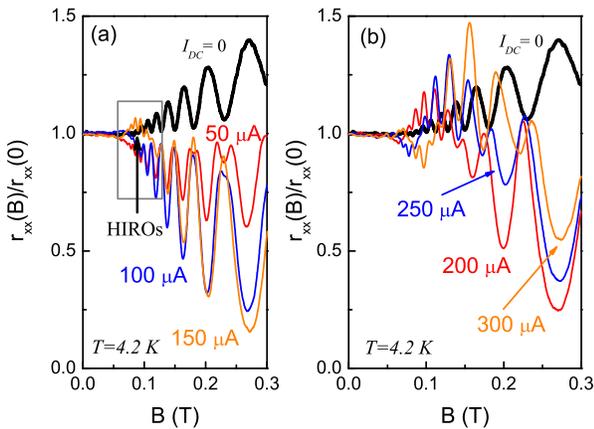}
\caption{\label{fig4} (Color online) HIRO's at a temperature of 4.2~K: a) for $I_{dc}$=0,
50, 100, and 150~$\mu$A and (b) for 200, 250 and 300~$\mu$A. The oscillating response 
to dc excitation weakens with increasing temperature. SdH oscillations are not present in the 
chosen range of magnetic field at this temperature.}
\end{figure}

In Fig. \ref{fig4} we show the normalized differential resistance as a function of $B$ 
at a temperature of 4.2~K. The behavior is basically the same as in Fig. \ref{fig2}. 
For $I_{DC}$=0, we observe MIS oscillations whereas SdH oscillations are not 
present anymore. The amplitudes of HIRO's are weakened at higher temperatures 
due to enhancement of inelastic electron-electron scattering and the corresponding 
decrease of the quantum lifetime of electrons \cite{20, 34}. 

Let us analyze the observed behavior in more detail. A theory which satisfactorily 
describes the origin of ZDRSs in our samples has been presented in Ref. \cite{33}. For 
the regions of magnetic field around 0.27 and 0.4 T, corresponding to inverted MIS 
peaks, the differential resistivity calculated within the assumption of homogeneous current 
appears to be negative if $T$ is low enough. The negative differential resistivity, according to 
Ref. \cite{11}, leads to current domains and to ZDRSs in the observed magnetoresistance as in 
the case of single-subband systems \cite{9,10}. The theory \cite{33} also explains the 
splitting of the maxima between inverted MIS peaks as a result of the influence of Landau 
quantization on inelastic electron-electron scattering in two-subband systems (see also Ref. 
\cite{31}). The effects of higher currents described above are consistent with this theory. 
In particular, the ZDRS is suppressed owing to heating of the 2DES and the related decrease of the 
quantum lifetime of electrons, which leads to a suppression of all kinds of magnetoresistance 
oscillations, especially for lower magnetic fields. The splitting disappears because at elevated 
currents ($\epsilon_{dc} \sim 1$) the influence of electron-electron scattering on the distribution 
function of electrons becomes inessential \cite{21}, this function is stabilized by the Hall 
field rather than by the inelastic scattering. We have confirmed these properties by a direct 
numerical calculation of the magnetoresistance. By including the short-range scattering potential 
into the theory (see the details below), we also checked the presence of the first HIRO peak 
near $B=0.1$ T at $I_{dc}=150$ $\mu$A. This peak coexists with the ZDRS around 0.4 T, in full 
accordance with our experimental data (see Fig. \ref{fig1}).

We now concentrate on theoretical calculations of magnetoresistance under strong dc excitation 
and weak magnetic fields, in the regime where HIROs are observed. In the description of electron 
transport, we assume elastic scattering of electrons by a random static potential of impurities 
or other inhomogeneities. If the region of magnetic field below 0.3~T we apply the approximation 
of overlapping Landau levels. The quantum contribution to resistivity in this regime is proportional 
to the square of the Dingle factor $d=\exp(-\pi/\tau_q \omega_c)$, where $\tau_q$ is the quantum 
lifetime of electrons. Under these conditions, non-linear magnetoresistance in single-subband systems
is described by the theory of Ref. \cite{21}. A generalization of this theory to two-subband systems 
is straightforward since the subband separation $\Delta$ is small compared to the Fermi energy 
$\varepsilon_F$, so the difference in Fermi velocities and scattering rates for different subbands 
can be neglected. In the regime of classically strong magnetic fields, the resistivity of the 
electron system is found from the following relation between the density of the applied current 
$j=I_{DC}/w$ and the longitudinal electric field $E_{||}=V_{xx}/l$:
\begin{eqnarray}
E_{||}=j \rho_0 \biggl\{ 1-d^2\left(1+\cos \frac{2 \pi \Delta}{\hbar \omega_c} \right) \nonumber \\ 
\times \left[ \frac{\partial^2 \gamma(\zeta)}{\partial \zeta^2} 
+ \frac{2 (\partial \gamma(\zeta)/\partial \zeta)^2}{\tau_{tr}/\tau_{in}+ 
\tau_{tr}/\tau_{q} - \gamma(\zeta)} \right] \biggr\},
\label{eq1} 
\end{eqnarray}
where $\rho_0=m/e^2n_s \tau_{tr}$ is the classical Drude resistivity, $\tau_{tr}$ is 
the transport time and $\tau_{in}$ is the inelastic scattering time. The dimensionless 
function $\gamma(\zeta)$ is given by the expression
\begin{equation}
\gamma(\zeta)= \int_{0}^{2 \pi} \frac{d \theta}{2 \pi} J_0[2 \zeta \sin(\theta/2)] 
\frac{\tau_{tr}}{\tau(\theta)}=  \sum_k \frac{\tau_{tr}}{\tau_k} J^2_k(\zeta),
\label{eq2} 
\end{equation}
where $J_k(x)$ are the Bessel functions and $1/\tau(\theta)$ is the angular-dependent 
scattering rate at the Fermi energy. The integral in Eq. (\ref{eq2}) describes the 
averaging over the scattering angle $\theta$. The second part of Eq. (\ref{eq2}) shows 
an equivalent representation of $\gamma(\zeta)$ in terms of angular harmonics of 
this scattering rate \cite{21}. The transport time $\tau_{tr}$ is defined in the usual 
way, as the inverse of the angular averaging of the factor $(1-\cos \theta)/\tau(\theta)$.
The dimensionless parameter $\zeta=\pi \epsilon_{dc}$ characterizes the Hall-field-induced 
tilt of Landau levels. One can write $\zeta$ through the current density as
\begin{equation}
\zeta=\sqrt{\frac{4 \pi^3 j^2}{e^2 n_s \omega_c^2}}.
\label{eq3} 
\end{equation}
The formal difference of the theory described by Eqs. (\ref{eq1})-(\ref{eq3}) from the single-subband 
theory of Ref. \cite{21} is given by the MIS oscillation factor $[1+\cos(2 \pi 
\Delta/\hbar \omega_c)]/2$ in the quantum contribution to the resistivity and by an extra 
factor of $1/\sqrt{2}$ in Eq. (3). 

\begin{figure}[ht]
\includegraphics[width=9cm]{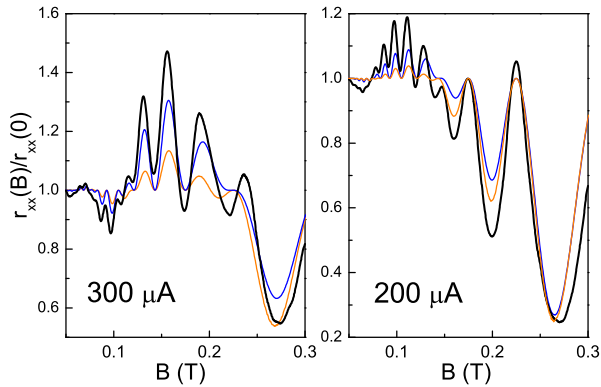}
\caption{\label{fig5} (Color online) Normalized differential resistance 
at $T=4.2$ K for $I_{dc}=300$ $\mu$A and $I_{dc}=200$ $\mu$A. Broad lines (black) are 
the experimental plots while narrow lines show the calculated magnetoresistance for the 
cases when the relative contribution of short-range scatterers into the transport 
scattering rate $1/\tau_{tr}$ is 50 \% and 90 \%. The amplitudes of the oscillations 
increase with increasing content of the short-range scatterers.}
\end{figure}

The calculation of the differential resistance from Eqs. (\ref{eq1})-(\ref{eq3}) is based upon 
the model of mixed disorder \cite{20,21,22}. We assume that the random static potential 
contains two components: the short-range and the long-range ones. The corresponding 
scattering rate is expressed as 
\begin{equation}
\frac{1}{\tau(\theta)}=\frac{1}{\tau_s}+\frac{1}{\tau_l(\theta)},
\end{equation}
where the angle-independent $\tau_s$ gives the contribution of the short-range 
potential into the total scattering rate. The long-range contribution is 
approximated by an exponential form $\tau_l^{-1} \propto \exp(-l_cq)$, where 
$\hbar q = 2 p_F \sin(\theta/2)$ is the transferred momentum, $p_F$ is the 
Fermi momentum, and $l_c$ is the correlation length of the potential.
For numerical calculations, we have used $\tau_{tr}$ found from the zero-field 
mobility and determined the quantum lifetime at $T=4.2$ K as $\tau_q = 6$ ps, based on 
the comparison between the measured and calculated magnetoresistance in the linear transport 
regime. The inelastic scattering time is estimated \cite{17} as $\tau_{in} \simeq \hbar 
\varepsilon_F/T_e^2$, where $T_e$ is the electron temperature. The appreciable 
heating of the electron gas by the applied field is obtained from the energy balance 
equation taking into account the energy loss due to electron-phonon interaction (for 
example, the current of 300 $\mu$A is found to heat electrons from $T=4.2$ K to 
$T_e  \simeq 4.8$ K). In Fig. 5 we plot the calculated normalized magnetoresistance 
for two different values of the applied current $I_{dc}$ and compare it with the 
experimental magnetoresistance. While for both currents we have a good agreement 
in the HIRO periodicity, the amplitudes of the oscillations expected from the theory 
are smaller than the experimental ones in the low-field part of the plot. 
Naturally, since the HIROs exist due to backscattering processes, their 
amplitudes increase with increasing content of the short-range scatterers.
This also leads to a better agreement between the theory and the experiment. 
Therefore, based on the large HIRO amplitudes observed, we may conclude that the 
amount of short-range scatterers in our samples is significant. 

\section{Influence of dc excitation on the MW-induced magnetoresistance}

In this section, we present experimental results of combined action of dc and ac 
excitations in our bilayer system. As it has been shown in Ref. \cite{28}, a ZRS 
occurs around $B=0.27$~T for 143~GHz at a temperature of 1.4~K. We focus on that 
particular frequency with the corresponding MW electric field of $E_{\omega} 
\simeq 4.2$~V/cm for 0~dB attenuation. As in previous studies \cite{25,27,28}, MIROs 
and ZRSs are not sensitive to the orientation of linear polarization. The immunity of 
MIROs and ZRSs to the sense of circular polarization has also been found in Ref. \cite{35} 
where the authors employed a quasioptical setup in which linear and any circular 
polarization can be produced \textit{in situ}.

\begin{figure}[ht]
\includegraphics[width=9cm]{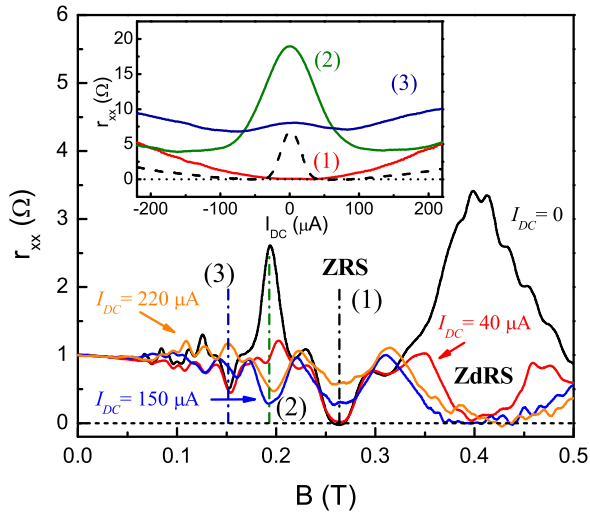}
\caption{\label{fig6} (Color online) Magnetoresistance under an AC excitation of 143~GHz
($I_{DC}$=0) and under combined ac and dc for several chosen DC currents of $I_{dc}$=40,
150 and $I_{DC}$=220~$\mu$A. The ZRS at 0.27~T is destroyed for 150 and 220~$\mu$A. A ZDRS
appears under MW excitation from the enhanced MIS oscillation at $B$=0.4~T for 
$I_{DC}>$40~$\mu$A. Inset: $r_{xx}$ under MW excitation for fixed $B$ for MIS peaks from
(1) to (3) as a function of $I_{dc}$ (dashed line without MW's, see also Fig. \ref{fig3}(a)).}
\end{figure}

In Fig. \ref{fig6} we plot the differential magnetoresistance for $I_{dc}$=0 and for different 
DC excitations. The zero-bias ($I_{dc}=0$) trace shows a ZRS developed from the inverted MIS 
peak at $B$=0.27~T [dashed-dotted line (1)], the other inverted MIS peaks at 0.15 [line (3)] 
and 0.1~T, as well as strongly enhanced MIS peaks around 0.19~T [line (2)] and 0.4~T. 
If we apply a dc bias starting with $I_{dc}$=40~$\mu$A, we obtain a strongly modified oscillation 
pattern: the MW-enhanced MIS peaks exhibit strong damping or even inversion, while the effect 
of $I_{dc}$ on the MW-inverted MIS peaks (including the one transformed into ZRS) is very weak. 
A further increase in dc excitation has two consequences: (i) the ZRS around $B=0.27$ T vanishes 
and (ii) a ZDRS develops under MW irradiation appears in a wide region around $B=0.4$~T. For the highest DC excitation 
of $I_{dc}=220$~$\mu$A, we also find well-developed HIRO's for $B < 0.2$~T. The dependence of 
the magnetoresistance on the dc current is non-monotonic for the traces (2) and (3).  
We have also carried out $B$ sweeps at different $I_{dc}$ up to 220~$\mu$A (not shown here) and
measurements of the resistance at constant magnetic field as a function of $I_{dc}$ (see inset of 
Fig. \ref{fig6}). For the ZRS at $B$=0.27~T, we find that a dc excitation leads to a vanishing ZRS
for $I_{dc}\geq50$~$\mu$A which corresponds to an electric field $E_{dc}>37$~V/m. At $B=0.41$~T, 
a ZDRSs develops under dc excitation develops for relatively low dc currents, $I_{dc} \simeq 28$~$\mu$A 
giving rise to $E_{DC}\simeq$39~V/m. 

A remarkable fact is that we see the ZDRSs in the same region 
of fields, $B \simeq 0.27$~T, regardless of the presence or absence of ac excitation, although with 
AC excitation we see the ZDRSs even in the linear response. As seen from the inset to Fig. \ref{fig6}, 
this state vanishes for $I_{dc}\geq50$~$\mu$A. For comparison, in the case of pure dc excitation the 
ZDRS appears at $I_{dc} \geq 40$~$\mu$A and vanishes at $I_{dc} \geq 80$~$\mu$A, see also 
Fig. \ref{fig3}(a). The observed behavior is imposed by the ability of both kinds of excitation 
to create in this region of magnetic fields a partially inverted electron energy distribution 
which leads to negative resistance. Thus, our observation is consistent with the domain model 
for the origin of both ZRSs and ZDRSs. 

The basic features of the observed behavior can be understood in terms of interplay of 
the effects of ac and dc excitations. The MW excitation either enhances or inverts MIS peaks 
(or groups of such peaks) in accordance with MIRO periodicity \cite{25}: the enhancement 
occurs around $\epsilon_{ac}=n+3/4$ (MIRO maxima) while the inversion corresponds to 
$\epsilon_{ac}=n+1/4$ (MIRO minima), $n$ being an integer. The DC excitation tends to invert 
all of the MIS peaks in the region of higher magnetic fields, where $\epsilon_{dc} < 1$. 
Thus, the MW-enhanced MIS peaks should be suppressed and eventually inverted by the DC 
excitation, while the peaks which are already inverted by MWs are expected to be little 
affected. However, the rapid suppression and inversion of the MW-enhanced MIS peaks already at a 
moderate DC excitation suggests that the interplay of ac and dc excitations is not reduced 
to a pure superposition of their effects. To make this point clear, we apply a simple 
theory valid at $\zeta^2 \ll 1$ and low temperatures, when the inelastic mechanism of 
non-linear response dominates, and use the approximation of overlapping LLs. The non-linear 
conductivity of a single-subband system in this case is given by Eq. (15) of Ref. \cite{17} and 
can be straightforwardly generalized for a system with two closely spaced subbands. Hence, 
the differential magnetoresistance in written in the form
\begin{eqnarray}
\frac{r_{xx}(B)}{r_{xx}(0)}=1+d^2 \left(1+\cos \frac{2 \pi \Delta}{\hbar \omega_c} \right) 
F(\epsilon_{ac},\epsilon_{dc}), ~~~~~ \\
F=\frac{(1+D)^2-10(1+D)Q-3Q^2+A(Q-1-D)}{(1+D+Q)^2}, \nonumber 
\end{eqnarray}
where $A =  2 \pi \epsilon_{ac} {\cal P}_{\omega} \sin 2 \pi \epsilon_{ac}$, 
$D= {\cal P}_{\omega} \sin^2 \pi \epsilon_{ac}$, ${\cal P}_{\omega} \propto 
\tau_{in} E^2_{\omega}$ is a dimensionless quantity proportional to MW power \cite{17}, 
and $Q = (\tau_{in}/2 \tau_{tr}) (\pi \epsilon_{dc})^2$ is a dimensionless quantity 
proportional to $I^2_{dc}$. In the MIRO minima and maxima, $A= \pm 2 \pi \epsilon_{ac} 
{\cal P}_{\omega}$, respectively, while $D={\cal P}_{\omega}/2$ in both cases (notice that 
$|A| \gg D$ for the relevant region $\omega > \omega_c$). For negative $A$ (MIRO maxima, 
enhanced MIS peaks), when all the $Q$-dependent terms in the numerator of $F$ have the same 
sign, the decrease of $r_{xx}$ with increasing $I_{dc}$ goes much faster than in the absence 
of MWs. Physically, this is related to a stronger effect of the dc field on the electron 
distribution if the latter is already modified by MWs. This explains the dramatic 
suppression of the magnetoresistance peaks by the dc excitation. For positive $A$ (MIRO 
minima, inverted MIS peaks), the MW term ($\propto A$) in $F$ considerably slows down 
the effect of $I_{dc}$ on magnetoresistance: the electron distribution is modified by 
MWs in such a way that it is only slightl affected by the dc field. However, in both cases
the increase in $I_{dc}$ leads to an additional heating of the 2DES. As a result, both 
$\tau_{in}$ and $\tau_q$ decrease, and so does the amplitude of oscillations. 
In Fig. \ref{fig6} we see this effect not only in the vanishing ZRS at 0.27 T [line (1)] 
but also in decreasing of the amplitude of the inverted peak at 0.19 T [line (2)]. We 
have estimated that the current $I_{dc}=150$ $\mu$A leads to an additional heating of 
the 2DES in our sample by approximately 1 K, which may explain the observed behavior. 
Below 0.2 T, Eq. (5) is no longer valid for the highest $I_{dc}$ used. 
In this region of fields, the non-linear magnetoresistance can not be simply described by
the ac and dc response of the 2D system (superposition) but rather as an interplay
(interference) of MIROs and HIROs.

\section{Conclusion}

We have studied magnetotransport in a high-quality bilayer electron system under a strong 
dc excitation. We have found well-developed HIROs which modulate the MIS oscillations 
in our bilayers. A theoretical calculation of the non-linear magnetoresistance gives a good 
agreement with experiment concerning the shape and periodicity of the oscillation picture. 
The comparison of theory and experiment indicates a significant role of electron scattering 
by the short-range random potential in our samples. Furthermore, we show that HIROs in the 
region of low magnetic fields coexist with zero differential resistance at higher fields. 

We have also examined non-linear transport under both ac (microwave) and dc excitations. 
Even a moderate dc excitation strongly modifies the MW-induced resistance. This effect is consistent 
with a theoretical model predicting that the combined action of two kinds of excitations is 
not merely a superposition of ac and dc effects. It is found that for the chosen MW frequency 
both these excitations lead to ZDRSs in the same region of fields, around $B=0.27$ T. The 
properties of ZDRSs with and without MW excitation are in accordance with the domain model 
which explains the vanishing resistance as a result of instability of homogeneous current flow 
under conditions of negative differential resistivity. A better understanding of the interplay of ac and 
dc excitations is desirable and requires further experimental and theoretical work.

We acknowledge support from COFECUB-USP (Project number U$_{c}$ Ph 109/08), FAPESP and CNPq
(Brazilian agencies).

\end{document}